\documentclass[lettersize,journal]{IEEEtran}
\usepackage{graphicx}
\usepackage{tcolorbox}
\usepackage{subeqnarray}
\usepackage{bm}
\usepackage{amsfonts}
\usepackage{amsthm}
\usepackage{amssymb}
\usepackage{makecell}
\usepackage{calc}
\usepackage{color}
\usepackage{mathrsfs}
\usepackage{flushend}
\usepackage{url}
\usepackage{caption}
\usepackage{subfloat}
\usepackage{color}
\usepackage{textcomp}
\usepackage{stfloats}
\usepackage{amsmath,amsfonts}
\usepackage{algorithmic}
\usepackage{algorithm}
\usepackage{array}
\usepackage[caption=false,font=normalsize,labelfont=sf,textfont=sf]{subfig}
\usepackage{textcomp}
\usepackage{stfloats}
\usepackage{url}
\usepackage{verbatim}
\usepackage{graphicx}
\usepackage{cite}
\usepackage{multirow}
\usepackage[outdir=./]{epstopdf}
\usepackage{booktabs}

\begin{document}

\title{Integrating Sensing and Communications in 6G? Not Until It Is Secure to Do So}

\author{Nanchi Su,~\IEEEmembership{Member,~IEEE}, Fan Liu,~\IEEEmembership{Senior~Member,~IEEE}, Jiaqi Zou,~\IEEEmembership{Graduate Student Member,~IEEE}, Christos Masouros,~\IEEEmembership{Fellow,~IEEE}, George C. Alexandropoulos,~\IEEEmembership{Senior~Member,~IEEE}, Alain Mourad,~\IEEEmembership{Member,~IEEE}, Javier Lorca Hernando,~\IEEEmembership{Member,~IEEE}, Qinyu Zhang,~\IEEEmembership{Senior~Member,~IEEE}, Tse-Tin Chan,~\IEEEmembership{Member,~IEEE}


\thanks{N. Su is with the Guangdong Provincial Key Laboratory of Aerospace Communication and Networking Technology, Harbin Institute of Technology (Shenzhen), China, and with the Education University of Hong Kong, Hong Kong SAR, China; F. Liu is with Southeast University, China; J. Zou is with Tsinghua University, China; C. Masouros is with University College London, U.K.; Alain Mourad and Javier Lorca Hernando are with InterDigital, U.K.; G. C. Alexandropoulos is with the National and Kapodistrian University of Athens, Greece; Q. Zhang is with the Guangdong Provincial Key Laboratory of Aerospace Communication and Networking Technology, Harbin Institute of Technology, China, and also with Peng Cheng Laboratory, China; T.-T. Chan is with the Education University of Hong Kong, Hong Kong SAR, China.}

}


\maketitle

\begin{abstract}
Integrated Sensing and Communication (ISAC) is emerging as a cornerstone technology for forthcoming 6G systems, significantly improving spectrum and energy efficiency. However, the commercial viability of ISAC hinges on addressing critical challenges surrounding security, privacy, and trustworthiness. These challenges necessitate an end-to-end framework to safeguards both communication data and sensing information, particularly in ultra-low-latency and highly connected environments. Conventional solutions, such as encryption and key management, often fall short when confronted with ISAC’s dual-functional nature. In this context, the physical layer plays a pivotal role: this article reviews emerging physical-layer strategies, including artificial noise (AN) injection, cooperative jamming, and constructive interference (CI), which enhance security by mitigating eavesdropping risks and safeguarding both communication data and sensing information. We further highlight the unique privacy issues that ISAC introduces to cellular networks and outline future research directions aimed at ensuring robust security and privacy for efficient ISAC deployment in 6G.
\end{abstract}

\section{Introduction}
\IEEEPARstart{T}{he} rapid evolution of communication systems toward 6G networks promises to deliver unprecedented connectivity, enabling a new era of intelligent, integrated services across diverse sectors, from healthcare to autonomous transportation. One of the core innovations expected in 6G is the integration of Sensing and Communications (ISAC), wherein communication networks seamlessly incorporate sensing capabilities to enhance not only system performance but also the services and applications built on top of these networks. This convergence introduces a range of transformative applications that go beyond traditional communication, creating new possibilities in fields such as autonomous driving, smart cities, environmental monitoring, public safety, and industrial IoT.

However, these technological advances have led to a growing volume and complexity of data exchange among devices, users, services, and network infrastructure in 6G ISAC systems, raising significant security and privacy concerns. With sensitive personal and operational data being generated and processed on an unprecedented scale, ensuring the confidentiality, integrity, and availability of this information has become critical. In this context, recent global developments in data protection regulations highlight the growing emphasis on security in digital ecosystems. For instance, while the European Data Act established clear and fair rules for accessing and using data within the European data economy, the European Union's General Data Protection Regulation (GDPR) continues to set the standard for privacy, imposing stringent rules on data processing and transfers. Similarly, China's Personal Information Protection Law (PIPL) enforces robust data protection practices, particularly concerning cross-border data flows, while California's Privacy Rights Act (CPRA) strengthens consumer rights to control sensitive personal information \cite{determann2020california}.

ISAC, along with its wide-ranging benefits, brings entirely new security vulnerabilities. First, as illustrated in Fig.~\ref{Security and privacy threats in ISAC systems.}, a critical challenge unique to ISAC arises when the sensing targets are non-cooperative or potentially malicious. In such cases, illuminating the targets with high-power information-carrying ISAC signals creates the opportunity for the targets to eavesdrop on the confidential data in the ISAC signal (Fig.~\ref{Security and privacy threats in ISAC systems.}(a)). More critically, the new sensing functionality of the ISAC infrastructure makes private information such as target location, size, shape, and mobility vulnerable to eavesdropping from unauthorized sensors (Fig.~\ref{Security and privacy threats in ISAC systems.}(b)). Current communication security schemes, such as end-to-end encryption, secure key management, and traditional authentication protocols, have been effective in safeguarding data in conventional networks. However, these methods encounter critical limitations with the integration of sensing functionalities. Traditional encryption techniques introduce latency due to the computational overhead associated with encryption and decryption processes, which is incompatible with the ultra-low latency (ULL) requirements of 6G applications. While data itself can be encrypted, one can still infer critical Layer1-Layer2 transmission parameters such as power, carrier frequency, and bandwidth, which can jeopardize the security of the data links. Additionally, current methods primarily focus on securing communication channels, while overlooking the new complexities imposed by their sensing counterparts, where sensitive sensory data is at higher risk of exposure. ISAC technologies cannot be deployed commercially until challenges are alleviated, and the ISAC systems are made secure for their users. To meet the demands of future networks, security frameworks must be highly adaptive, while simultaneously protecting the communication data and sensing privacy.

The aforementioned limitations have spurred a growing interest in physical layer security (PLS) research as an essential component of the security and privacy framework for ISAC networks. PLS can provide real-time security without the time delay resulting from the computational overhead, and it is inherently scalable as it does not rely on complex key management systems. Notably, the new sensing functionality brings new opportunities for ISAC security. With the enabling role of sensing in providing environment awareness, ISAC systems can estimate unauthorized sensing channels (i.e., channels used for illicit reception of communication data). This approach addresses the challenge of acquiring eavesdropping channel information~\cite{10143983}, which has been a significant barrier in applying PLS in practice in the past. Additionally, the dual-functional waveform provides new avenues of privacy protection, where novel techniques can ensure that adversaries are unable to extract sensitive information through side-channel attacks or unintended signal leakage. 

These integrated approaches can fundamentally reshape the security landscape of next-generation wireless networks, making ISAC systems not only more efficient but also inherently more secure. This article delves into these advancements, offering a comprehensive review of the advancing PLS and privacy-preserving methods for future ISAC networks.

\section{Security and Privacy Threats in ISAC}
\subsection{ISAC Basics}
Given that ISAC has been extensively reviewed in the existing literature, this subsection provides a concise introduction, focusing on three key aspects: the physical layer, the network layer, and joint hardware design.

1) Physical layer techniques: 
Waveform designs for ISAC systems have been extensively explored and overviewed in the existing literature, encompassing communication-centric, sensing-centric, and joint waveform designs. These methodologies have been thoroughly reviewed in \cite{lu2024integrated}, demonstrating that the dual-functional waveforms have enabled complex implementations in vehicle-to-everything (V2X) communications, unmanned aerial vehicles (UAVs), smart cities, etc. Recent work has also extended the study to the fundamental limits of ISAC, discussing fundamental tradeoffs between the Cram\'er-Rao Bound (CRB) and capacity distortion (C-D) \cite{xiong2024torch}. 

2) Network layer schemes: ISAC research at the network level focuses on enhancing cooperation and synergy between S\&C functionalities across multiple cellular sites while also providing inherent mechanisms to collect, process, and expose sensing data both within the network and to external applications. Key advancements in network-level ISAC include the development of cooperative multi-cell strategies building on communication-only methodologies such as coordinated multi-point (CoMP). Tailoring these for ISAC expands coverage and enhances the precision of both communication and sensing capabilities \cite{meng2024cooperative}. Moreover, integrating new sensing sources into the cellular infrastructure brings additional challenges in data fusion, real-time analytics, resource management, and security. Addressing these complexities, along with considerations such as signaling overhead and synchronization, calls for innovative network architectures and protocols that balance the performance gains of ISAC against the increased overhead, while ensuring that sensing data is efficiently gathered, processed, and accessible for diverse use cases \cite{strinati2025toward}.

3) Hardware architectures for ISAC:
The hardware design for ISAC systems emphasizes advanced radio frequency (RF) front-end architectures and antenna designs tailored for dual-functionality in S\&C. This involves integrating wideband transceivers capable of supporting various signal types and modulation schemes to facilitate both high-resolution sensing and efficient data communication. Moreover, the design also considers low-latency signal processing hardware to manage the real-time requirements of ISAC systems, ensuring that the hardware can swiftly switch between S\&C modes without performance degradation. These developments are crucial in pushing the boundaries of ISAC technology toward practical deployment in future communication networks.

In light of the above, ISAC systems have garnered considerable attention as a crucial enabler for future wireless networks. However, the deployment of ISAC hinges on ensuring robust security and privacy for both the communication data of users and the sensing information of targets. The potential data security and privacy threats in ISAC are elaborated in the following. 
\begin{figure}
  \centering
  \includegraphics[width=0.5\textwidth]{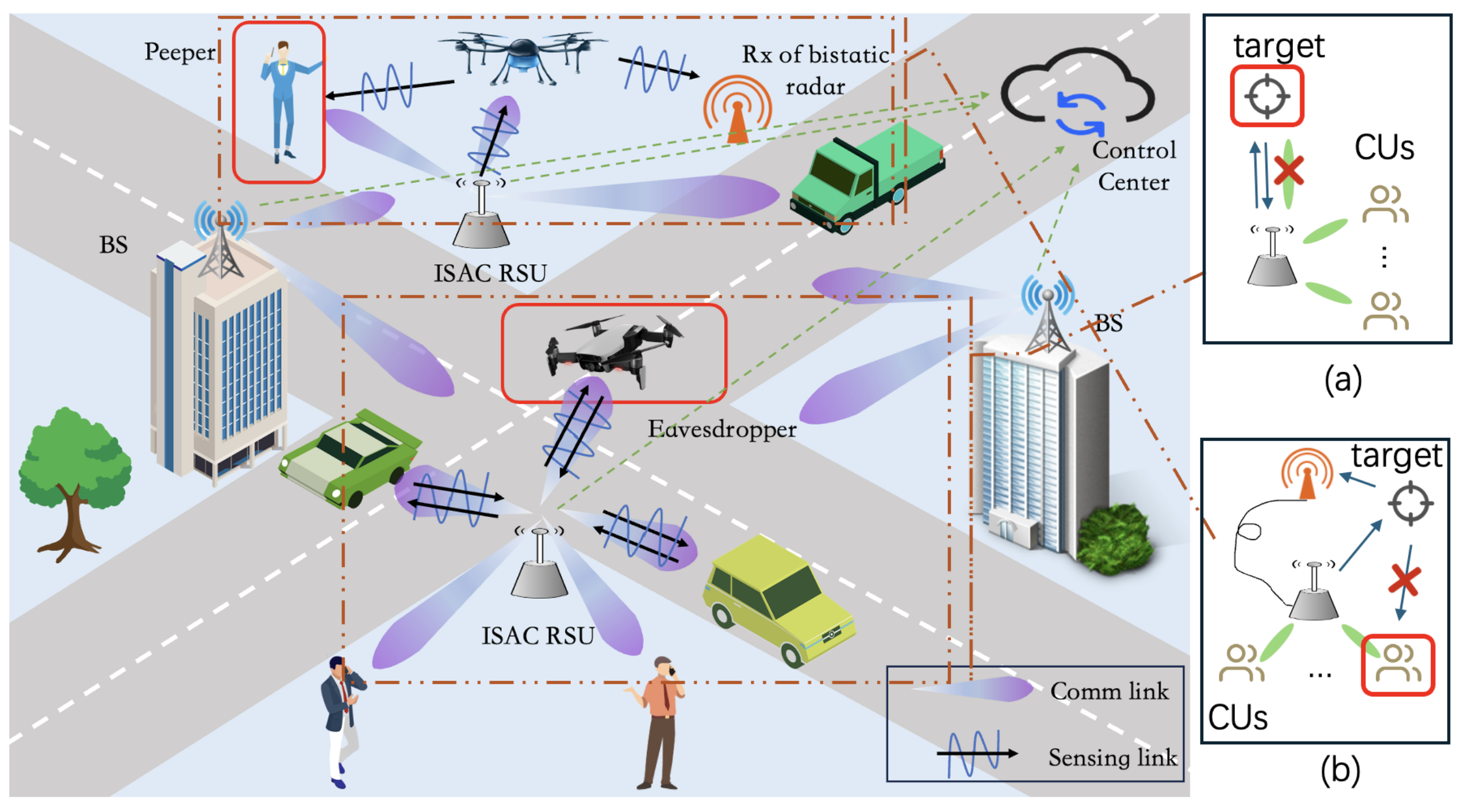}
  \captionsetup{font={footnotesize}}
  \caption{Data security and sensing privacy threats caused by unauthorized receivers in ISAC systems: (a) Data security, (b) Sensing privacy.} 
  \label{Security and privacy threats in ISAC systems.}
\end{figure}
\subsection{Data Security Threats in ISAC}
The platform integration of the ISAC significantly heightens security vulnerabilities. The security of communication data in ISAC systems can be comprehensively analyzed by examining threats at both the physical and network layers, which can be specified as follows.

\textbf{MAC layer and network layer:} 
    The shared data pathways between these dual functions increase the risk of routing integrity compromises, allowing attackers to exploit sensing-derived information for spoofing or impersonation attacks. This convergence also amplifies the likelihood of man-in-the-middle (MITM) attacks, where adversaries intercept or manipulate communication streams by leveraging sensing-related feedback or configurations. Furthermore, the joint management of S\&C data complicates system configuration, increasing the risk of errors that can lead to data breaches and unauthorized access. These threats underscore the critical need for security mechanisms that address the dual-purpose nature of ISAC, ensuring both sensing-derived data and traditional communication streams are protected in complex, interconnected environments.
    
 \textbf{Physical layer:} 
    Fundamentally, PLS needs to be jointly designed with sensing operation, leading to new joint PLS and sensing metrics, methodologies, and entirely new performance tradeoffs. More importantly, different from the nature of communication-only scenarios, at the physical layer of ISAC systems, the transmission of electromagnetic waves presents a unique challenge. Unlike radar-only systems, the probing signal contains communication information, making it prone to potentially malicious targets acting as eavesdroppers to extract critical information. This happens because ISAC may employ the same directional beam for dual purposes: to facilitate communication with users and to illuminate the target for sensing functionality. As the beam is directed explicitly toward the target to capture reflected or scattered signals essential for sensing, the target, strategically located within the beam's path, may receive all transmitted data embedded within this beam. Therefore, ISAC inherently exposes vital communication data to potential interception. Classical PLS, such as secure beamforming and null-steering to the target, do not apply, as they would jeopardize the sensing functionality. This necessitates the conception of entirely new PLS approaches tailored for ISAC.


\begin{table*}[t!]
    \centering
    \renewcommand{\arraystretch}{1.3}
    \caption{Summary of ISAC Data Security and Sensing Privacy Designs}
    \scriptsize
    \begin{tabular}{p{1.5cm}p{2cm}p{4.5cm}p{4cm}p{4cm}}
        \toprule
        \textbf{Category} & \textbf{Technique} & \textbf{Design Principles} & \textbf{Pros} & \textbf{Cons} \\
        \midrule
        \multirow{2}{*}{\shortstack{ISAC Data \\Security Design} }
        & AN Injection \cite{su2020secure}& Introduces artificially designed noise to degrade eavesdroppers' SNR while preserving legitimate communication. & Effective against eavesdroppers, adaptable to ISAC. & Requires accurate CSI for optimal interference placement. \\
        & CI \& DI \cite{su2022secure}& CI enhances legitimate signal detection, while DI disrupts eavesdroppers' reception by steering signals into undecodable regions. & Energy-efficient alternative to AN, well-suited for mmWave ISAC due to angular dependency. & Requires precise angle estimation and channel state feedback. \\
        \midrule
        \multirow{3}{*}{\shortstack{ISAC Sensing \\Privacy Design} }
        & Sensing Privacy via Minimizing Mutual Information \cite{10587082} & Limits the amount of information leaked to unauthorized sensing entities by minimizing mutual information. & Preserves legitimate sensing while restricting unauthorized sensing. & Requires real-time interference management for effective control. \\
        & Transmitter Privacy Protection \cite{al2019mimo}& Prevents adversaries from extracting radar location information using waveform design and gradient-based precoding. & Prevents localization-based attacks and mitigates side-channel vulnerabilities. & May slightly reduce sensing accuracy in highly dynamic environments. No countermeasures to the training-based adversaries. \\
        & User Location Privacy \cite{roth2021localization, roth2021privacy}& Prevents adversarial localization through CSI leakage mitigation techniques, such as randomized precoding. & Randomization reduces tracking accuracy without significantly affecting legitimate communication. & Requires additional processing to avoid unnecessary degradation in system performance. \\
        \midrule
        \multirow{2}{*}{\shortstack{ISAC-Enabled\\Privacy and \\Security}} 
        & Sensing-Assisted PLS \cite{su2023sensing} & Uses ISAC’s sensing capability to infer eavesdropper locations and adjust security measures dynamically. & Enhances proactive security by leveraging real-time sensing. & Performance depends on sensing accuracy and environmental stability. \\
        & Sensing-Assisted Covert Communication \cite{10473676}& Uses sensing data to predict adversary behavior and design undetectable communication strategies. & Achieves high confidentiality for secure transmissions. & Requires sophisticated environmental modeling and tracking. \\
        \bottomrule
    \end{tabular}
    \label{tab:isac_security}
\end{table*}

\subsection{Sensing Privacy Threats in ISAC}
On top of the new data security challenges, the privacy for sensing is also jeopardized due to the introduction of the sensing functionality, now inherent in the cellular network signals, architecture, and hardware. When sensing signals emitted by the dual-function base station are reflected by a target and subsequently intercepted by unauthorized receivers, there is a substantial likelihood that these users can obtain target-related information. This scenario presents a significant risk of sensitive target information leakage. Importantly, higher layer encryption does not apply, as there is no data link to encrypt; it is merely the ability of a node to sense the environment in a passive manner, exploiting the enhanced sensing signaling of ISAC technologies.

In this subsection, we explore these sensing privacy risks in detail, highlighting the vulnerabilities associated with ISAC waveforms and the subsequent threats to sensing privacy.

\textbf{Transmitter privacy:} 
    In ISAC systems, the specifically designed signal to mitigate mutual interference between S\&C contains implicit information about the transmitter, such as its geometrical information. This leads to the exposure of position and operational parameters to unauthorized receivers, thereby posing potential risks to transmitter privacy. Such threats have been highlighted in \cite{dimas2017spectrum}, which illustrates how the information embedded within a precoder could be exploited by an adversary to estimate the radar's location.

\textbf{Target confidentiality}:
    The enhanced sensing capabilities for ISAC signals open up opportunities for independent passive sensing from unauthorized nodes. Critically, this offers enhanced sensing eavesdropping performance since ISAC transmission is tuned for sensing. Sensing eavesdroppers can be both nodes external to the ISAC networks and nodes of the network that are not authorized for sensing. An example would be a communication user (CU) that is subscribed to communication services but not to sensing services.

\textbf{Imaging and environment privacy:} 
    In existing and emerging network technologies, significant risks to location and environment privacy arise from exploiting unencrypted control signals, such as channel state information (CSI). Passive eavesdropping attacks leverage these signals, especially the feedback of precoder indices, to deduce user positions \cite{roth2021localization}. Attackers construct unauthorized environmental maps by gathering precoder feedback from multiple positions within the cell and matching these maps with intercepted feedback from the victim. This method, utilizing the Type-I single-panel codebook standardized in 3GPP, exploits the direct relationship between precoder selection and spatial channel characteristics \cite{roth2021privacy}. Such techniques achieve high localization accuracy, particularly in millimeter-wave networks, where finer feedback granularity enables precise positioning.

In light of the above, ISAC systems are susceptible to various data security and sensing privacy threats, such as eavesdropping and adversarial sensing. To address these challenges, a range of countermeasures has been proposed, as summarized in Table I. The following sections provide a detailed analysis of these ISAC data security and sensing privacy solutions.

\begin{figure*}[t!]
    \centering
    \noindent 
    \begin{minipage}{0.55\textwidth}
        \centering
        \includegraphics[width=\linewidth]{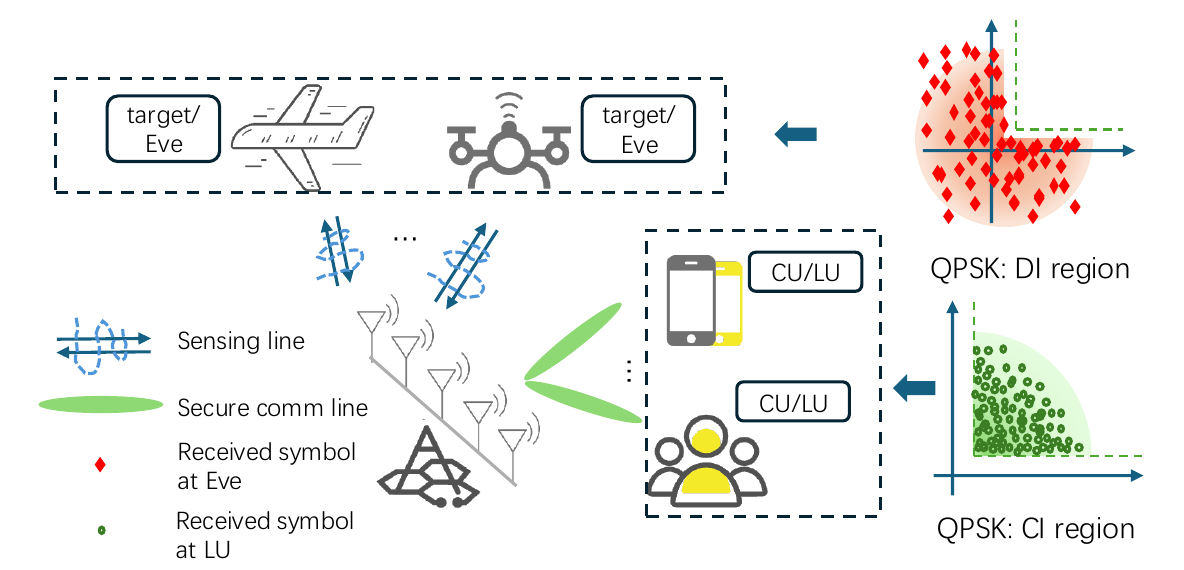} 
        \caption*{(a)} 
    \end{minipage}
    \hfill
    \begin{minipage}{0.2\textwidth}
        \centering
        \includegraphics[width=\linewidth]{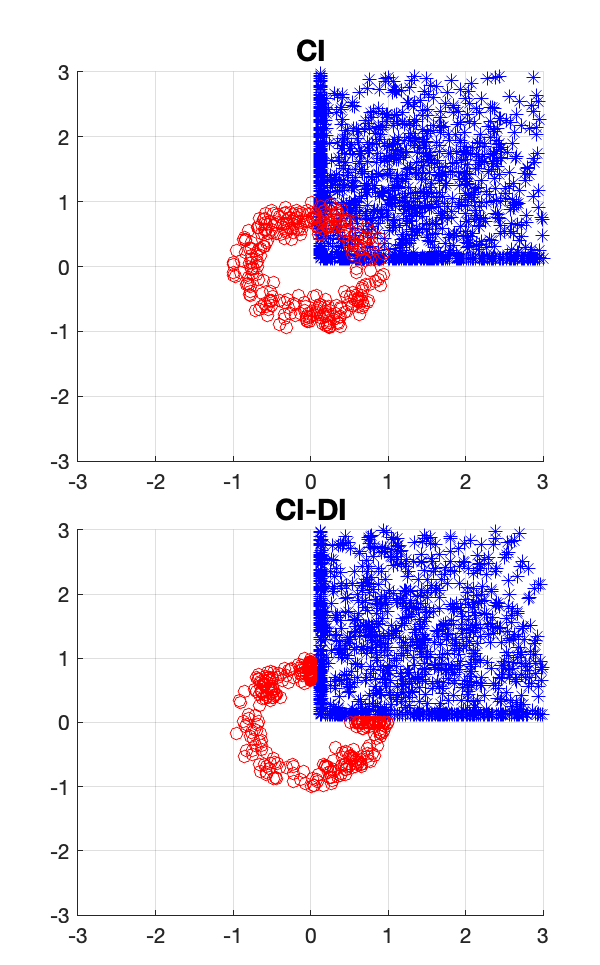} 
        \caption*{(b)} 
    \end{minipage}
    \hfill
    \begin{minipage}{0.2\textwidth}
        \centering
        \includegraphics[width=\linewidth]{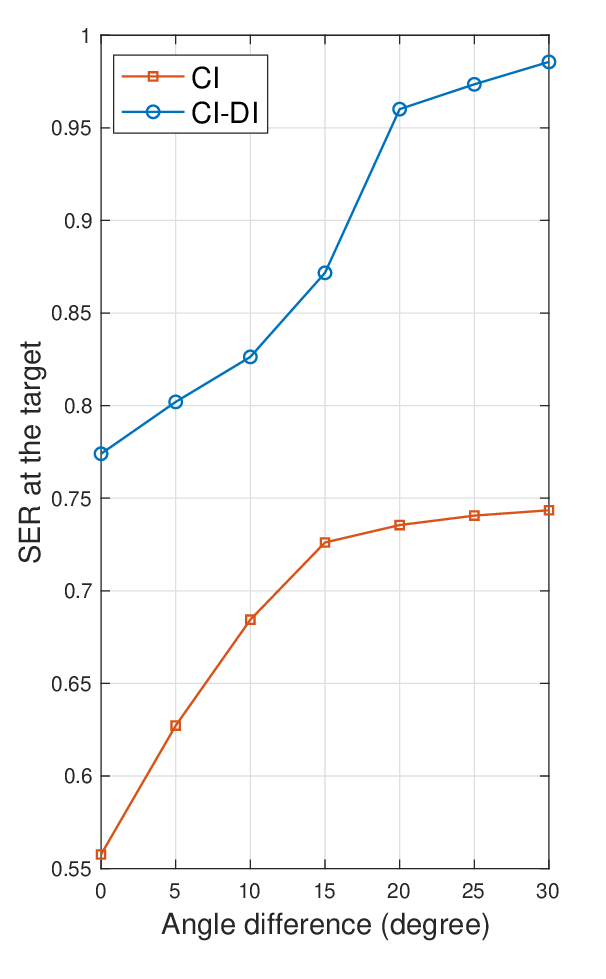} 
        \caption*{(c) } 
    \end{minipage}
    \caption{Communication data security in ISAC systems: CI-DI technique, where each CU is regarded as a legitimate user (LU), and the target is regarded as a potential eavesdropper (Eve). (a) System model: the dual-functional AP ensures communication data security by deploying the symbol-level precoder. (b) Constellations for CI technique (up) and CI-DI technique (down). (c) The symbol error rate (SER) of deploying the CI and the CI-DI techniques when the S\&C are correlated, where the correlation is depicted by the angle difference in mmWave frequency.} 
    \label{Communication data security}
\end{figure*}

\section{ISAC Data Security Solutions}
\subsection{ISAC-Tailored Artificial Noise (AN) Design}
AN is an interference-based technique that selectively degrades eavesdroppers’ reception while legitimate receivers remain unaffected. By reducing the SNR at unauthorized users, AN prevents malicious interception without compromising authorized data transfer. AN-aided methods have proven effective in mitigating security vulnerabilities in ISAC systems, particularly against malicious targets attempting to eavesdrop on confidential information. By injecting carefully designed interference into legitimate communication channels, AN techniques degrade the signal-to-noise ratio (SNR) at potential eavesdroppers while ensuring reliable communication for authorized users, as illustrated in Fig.~\ref{Security and privacy threats in ISAC systems.}(b). Recent advancements, such as the optimization framework proposed in \cite{su2020secure}, employ dynamic AN power and spatial distribution adjustments based on CSI, achieving an optimal tradeoff between data security, communication performance, and sensing accuracy.

Nevertheless, the dependency on precise CSI poses challenges in practical deployments. To overcome this challenge, robust optimization approaches and statistical CSI models have been developed to ensure consistent security performance despite CSI uncertainties. These innovations underline the significance of AN techniques as a cornerstone for secure and efficient communication in next-generation ISAC systems.

\subsection{Constructive-Destructive Interference}
Constructive Interference (CI) strengthens desired signals at legitimate receivers, while Destructive Interference (DI) deliberately weakens signals in unwanted directions. Directional Modulation (DM) leverages CI and DI to ensure the QoS for authorized users, whereas severely degraded signals for unauthorized receivers. 

As a step beyond AN, DM presents significant advantages, making it a promising PLS solution for 6G communication systems. Unlike conventional beamforming, which relies on expensive and power-intensive RF chains and digital-to-analog converters (DACs), DM-based schemes offer cost and energy efficiencies crucial for sustainable and scalable network operations. However, integrating DM into an ISAC framework introduces the additional challenge of jointly optimizing both communication and sensing performance, a requirement absent in classical DM systems.

DM leverages CI by ensuring that received signals at legitimate users are pushed away from detection thresholds rather than aligning precisely with intended symbols, as illustrated in Fig.~\ref{Communication data security}. Conversely, DI can be utilized to ensure that information-carrying signals used to illuminate targets are protected against potential eavesdropping \cite{su2022secure}. This technique enhances PLS by limiting the ability of eavesdroppers to decode signals. In the ISAC context, this design must also accommodate sensing performance, ensuring that sensing tasks are not compromised while securing communication data.

A key challenge of CI-enabled PLS is the correlation between legitimate and wiretap channels, which constrains secure transmission rates. This issue is further pronounced in ISAC scenarios, where sensing and communication functionalities must be jointly optimized. In mmWave ISAC systems, the correlation between communication and sensing channels depends on angular differences, with the communication channel varying based on the CUs' angles. This necessitates careful joint optimization of PLS and sensing performance, considering angular correlation. Numerical results in Fig.~\ref{Communication data security}(b) and (c) verify the effectiveness of the DI technique even when CUs and targets have small angular separations, demonstrating the feasibility of DM-based ISAC while addressing secure communication and robust sensing challenges.

\section{ISAC Sesing Privacy Solutions}
\subsection{Beamforming for Sensing Privacy}
\begin{figure}[t!]
    \centering
    \noindent 
    \begin{minipage}{0.23\textwidth}
        \centering
        \includegraphics[width=\linewidth]{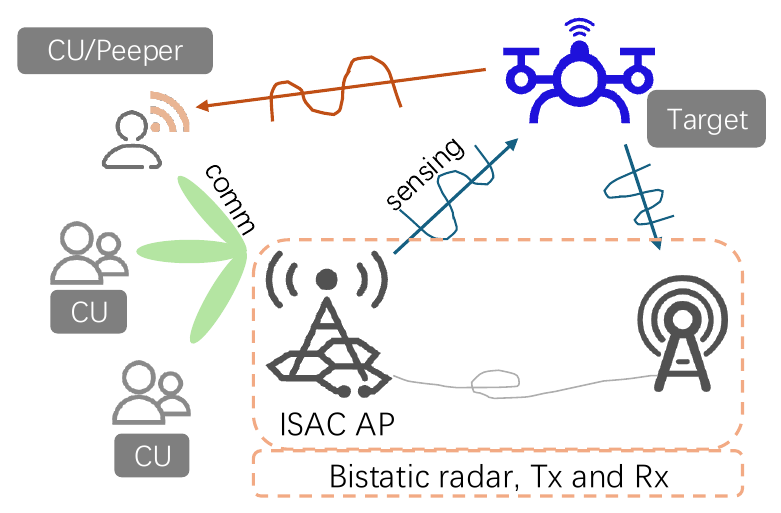} 
        \caption*{(a)} 
        \label{fig:subfig2} 
    \end{minipage}
    \begin{minipage}{0.25\textwidth}
        \centering
        \includegraphics[width=\linewidth]{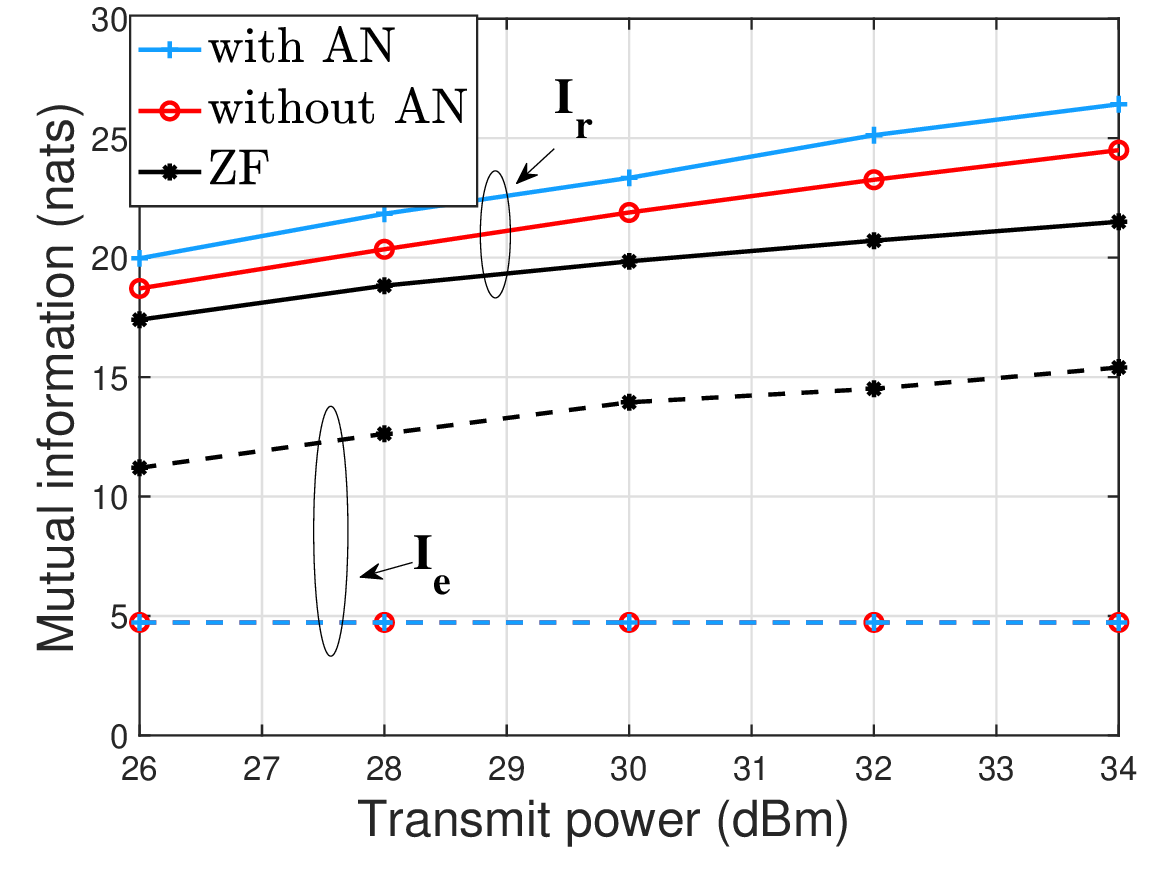} 
        \caption*{(b)} 
        \label{fig:subfig3} 
    \end{minipage}
    \caption{Sensing privacy: parameter leakage risks of the sensing targets. (a) System model: In a bistatic ISAC system, a central BS transmits dual-functional waveforms intended for both a legitimate radar receiver and $K$ single-antenna CUs. Meanwhile, one of the CUs, serving as an unauthorized peeper/Eve, has perfect knowledge of the transmitted signals. (b) The mutual information gap while deploying the ZF technique, with and without AN designs, where $I_r$ is the MI of the legitimate radar receiver and $I_e$ is the MI of the peeper.} 
    \label{parameter leakage risks}
\end{figure}

A particularly critical scenario arises when a CU subscribes only to communication services but attempts to exploit ISAC signals for unauthorized sensing activities. As illustrated in Fig.~\ref{parameter leakage risks}(a), CUs could potentially use the transmitted waveforms to gather information about surrounding targets or the environment, despite not being granted access to sensing services. This necessitates dedicated mechanisms to restrict their ability to perform unauthorized sensing.

To mitigate this security threat, the authors in \cite{10587082} propose an innovative beamforming design that maximizes mutual information (MI) for legitimate radar sensing receivers while limiting the MI accessible to unauthorized sensors, including eavesdropping CUs. Unlike traditional AN-aided communication security schemes, which focus solely on protecting data transmission, the proposed approach leverages AN to enhance sensing privacy. This dual-purpose strategy not only improves the sensing capabilities of legitimate receivers but also actively degrades the sensing performance of unauthorized users. The effectiveness of these methods is validated through simulations, as shown in Fig.~\ref{parameter leakage risks}(b), which demonstrate that AN significantly bolsters sensing security by increasing the MI gap between legitimate and unauthorized entities, all while maintaining the quality of service for CUs.

\subsection{Transmitter Privacy}
\begin{figure}[t!]
    \centering
        \noindent 
        \begin{minipage}{.45\columnwidth}
            \centering
            \includegraphics[width=\linewidth]{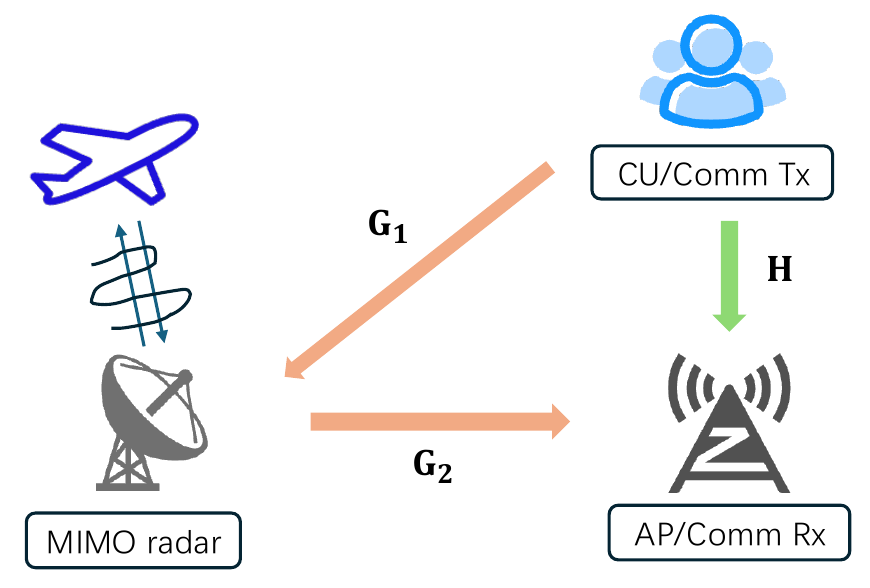} 
            \caption*{(a) } 
        \end{minipage}
        \begin{minipage}{.5\columnwidth}
            \centering
            \includegraphics[width=\linewidth]{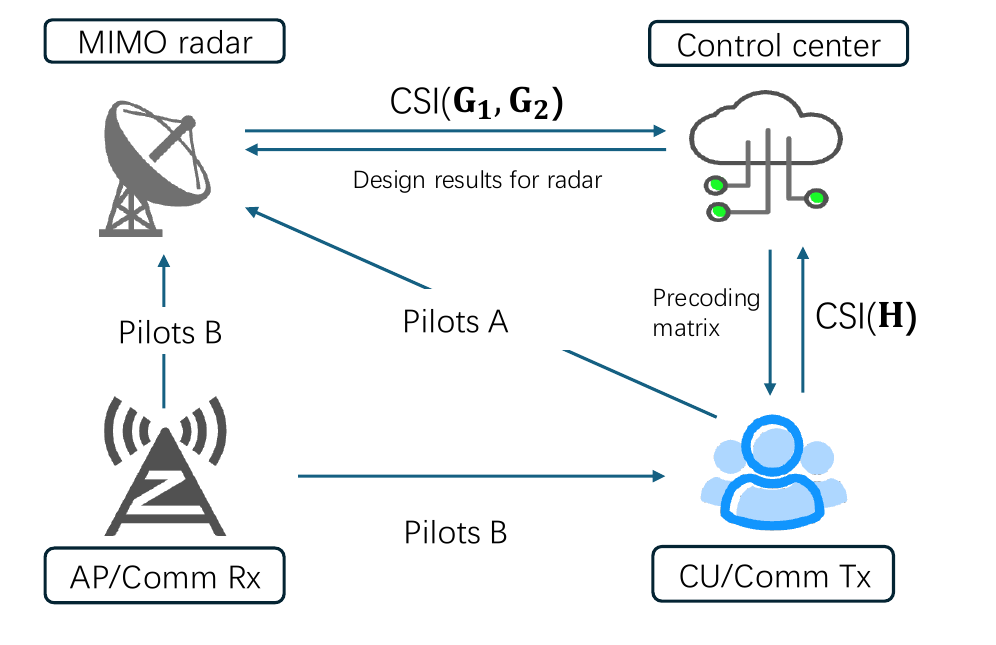} 
            \caption*{(b) } 
        \end{minipage}
    \caption{Sensing privacy: radar information privacy. (a) Spectrum sharing between co-located MIMO radar and communication systems; (b)Spectrum sharing architecture aided by a control center. $\mathbf{G}_1$,  $\mathbf{G}_2$, $\mathbf{H}$ denote channels: $\mathbf{G}_1$ from the radar to the CU, $\mathbf{G}_2$ from the radar to the access point (AP)/Comm receiver, and $\mathbf{H}$ from the CU to the AP/Comm receiver.} 
    \label{radar information privacy}
\end{figure}

Recall Section II-C, radar privacy can be compromised in both non-training-based and training-based frameworks \cite{dimas2017spectrum}, in spectrum sharing ISAC scenarios. To bridge this gap, authors in \cite{al2019mimo} proposed leveraging channel characteristics as a defense mechanism against radar privacy interception. Specifically, in Fig.~\ref{radar information privacy}, interference from communication devices to radar can be mitigated via optimization-based precoding. The communication precoder is designed to minimize interference at the radar while meeting power and rate constraints. Meanwhile, an adversary with knowledge of the waveform covariance matrix can estimate radar angles by scanning for the minimum power consumption.

To counter radar privacy leakage, a privacy-preserving precoder design enforces a positive gradient for the term associated with radar angles in the optimization problem. This prevents adversaries from exploiting angular scans to extract radar information, thereby safeguarding radar privacy in ISAC systems.

\subsection{User Location Privacy}
To mitigate location leakage risks from precoder index eavesdropping, randomizing the precoder index selection has been proposed as an effective countermeasure. Rather than always selecting the optimal precoder, users randomly choose from several top-performing precoders, introducing unpredictability into the transmission process. This strategy significantly reduces the accuracy of location inference attacks with minimal impact on overall communication performance. Such measures are critical for enhancing user privacy in 5G networks, where the precise nature of localization techniques poses substantial privacy risks \cite{roth2021localization}.

Balancing user location privacy with high-precision localization is a complex challenge. \cite{roth2021privacy} proposes solutions such as the integration of Reconfigurable Intelligent Surfaces (RISs) and the strategic manipulation of precoder feedback. RISs dynamically optimize signal paths during localization, restricting the precision of location data to legitimate users while obscuring it from potential adversaries. By focusing signals in specific directions, RIS configurations enhance the desired signal quality for user equipment (UE) while distorting or nullifying locational clues, thereby reducing the risk of location leakage. 


\begin{figure*}[t!]
    \centering
    \noindent 
    \begin{minipage}{.8\textwidth}
        \centering
        \includegraphics[width=\linewidth, height = 0.17\textheight]{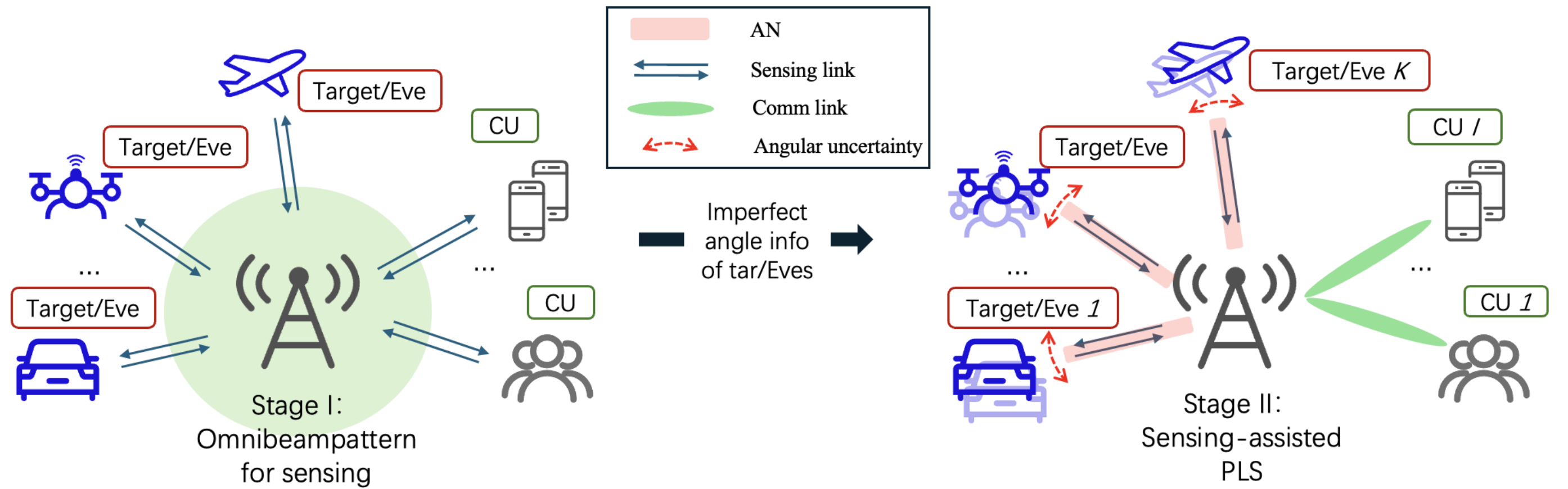} 
        \caption*{(a)} 
    \end{minipage}
    \hfill
    \begin{minipage}{0.3\textwidth}
        \centering
        \includegraphics[width=\linewidth, height = 0.18\textheight]{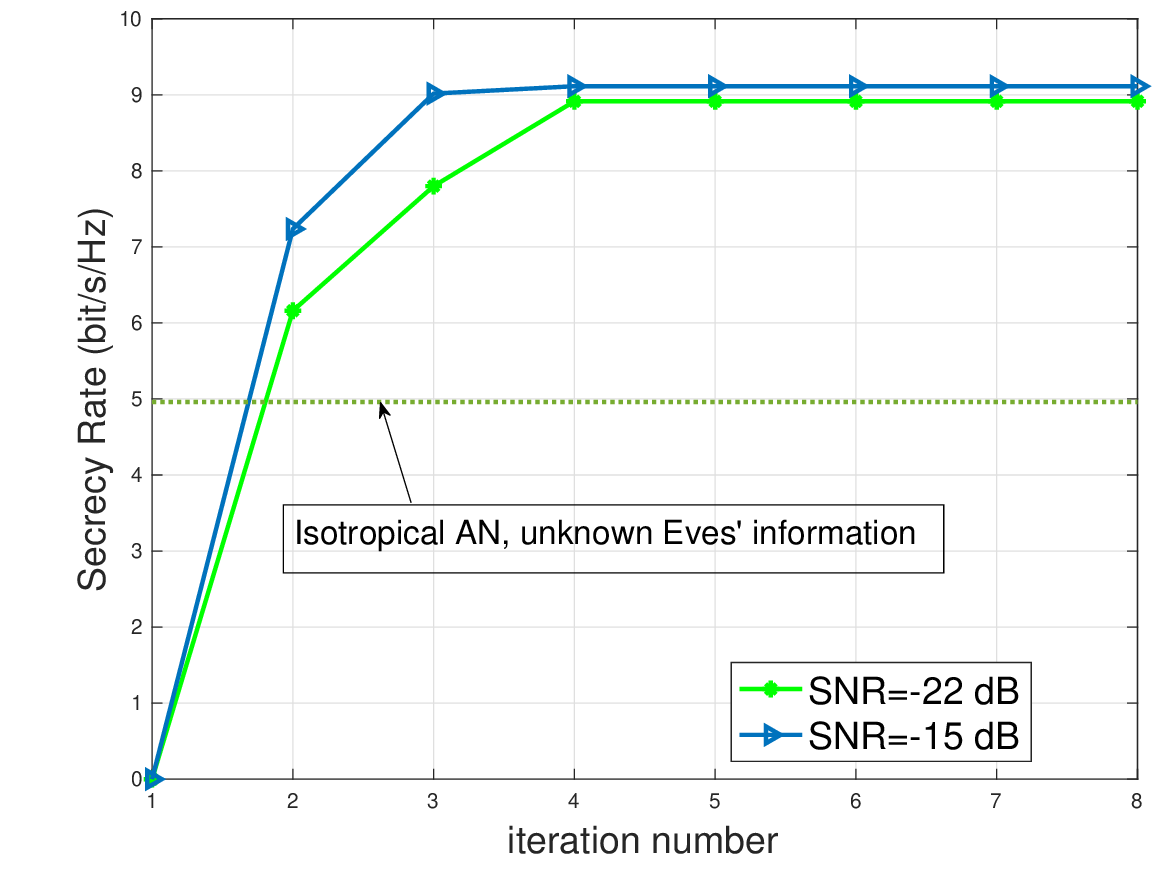} 
        \caption*{(b)} 
    \end{minipage}
    \hfill
    \begin{minipage}{0.3\textwidth}
        \centering
        \includegraphics[width=\linewidth, height = 0.18\textheight]{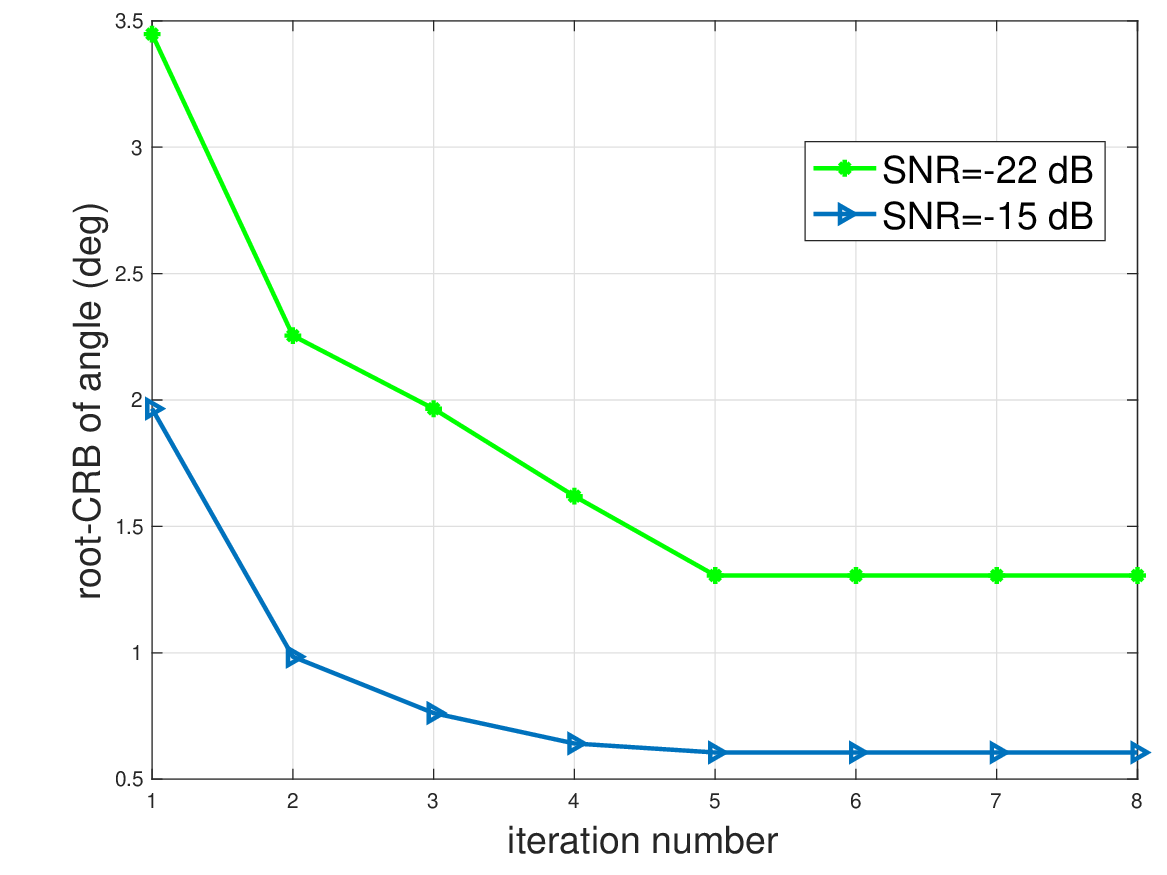} 
        \caption*{(c) } 
    \end{minipage}
    \hfill
    \begin{minipage}{0.3\textwidth}
        \centering
        \includegraphics[width=\linewidth, height = 0.18\textheight]{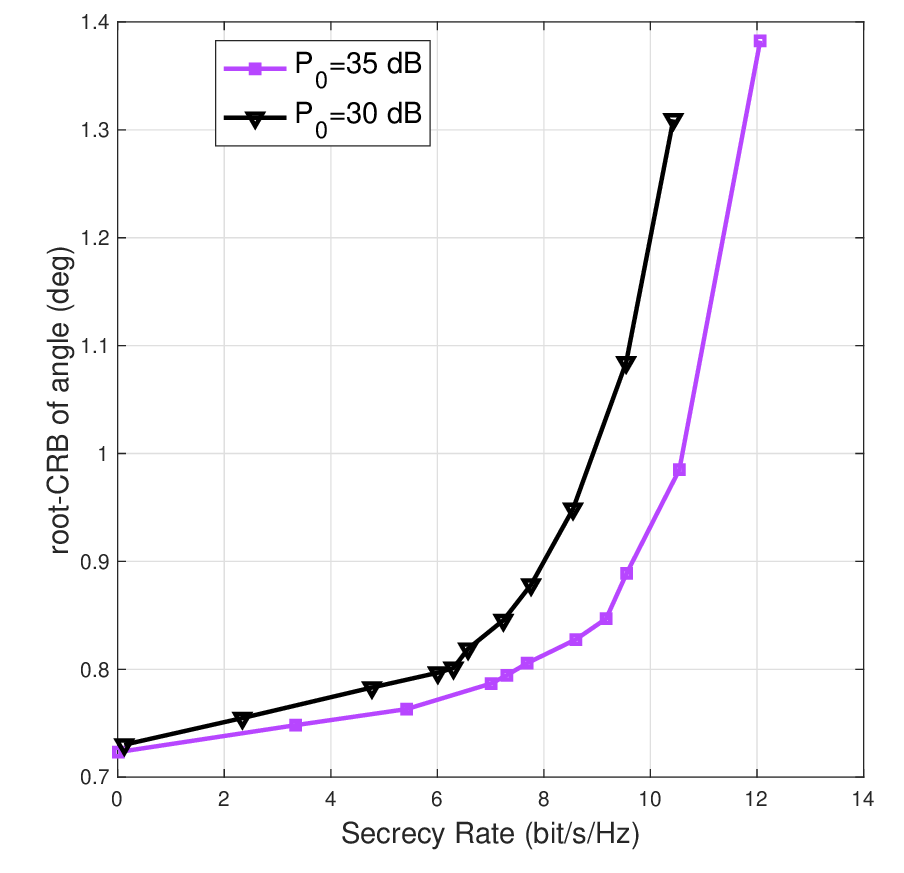} 
        \caption*{(d)} 
    \end{minipage}
    \caption{The two-stage sensing-assisted PLS technique in ISAC and the corresponding performance analysis. (a) The two-stage sensing-assisted PLS; (b) Beampattern convergence: single-target scenario; (c) Beampattern convergence: multi-target scenario; (d) Tradeoff between the PLS and the sensing performance. } 
    \label{two-stage}
\end{figure*}

\section{ISAC Enabled Privacy and Security}
\subsection{Sensing- Assisted PLS}
Leveraging the inherent advantages of ISAC, the BS can extract the amplitude and phase of each target within the line-of-sight (LoS) distance through reflected echoes. When this is used to sense potential eavesdroppers, this enables the acquisition of wiretap channel information. By recognizing that the angle of the target/Eve cannot always be estimated accurately, one effective design of secure dual-functional transmission is to maximize the secrecy rate, while shaping the radar beampattern such that the main beam may cover all possible angles of each target so as to ensure the sensing performance. This beamformer design effectively ensures the sensing performance and communication data security, where the detailed discussion can be found in \cite{su2020secure}. 

A two-stage sensing-assisted PLS technique was proposed in \cite{su2023sensing} to alleviate the dependence on the knowledge of the channel conditions, as shown in Fig.~\ref{two-stage}. In this case, the sensing performance is measured by the Cram{\'e}r-Rao Lower Bound (CRLB), and the PLS is measured by the secrecy rate. The unique insight of this work is that even without \textit{a priori} knowledge of the wiretap channel, the sensing functionality can still obtain eavesdropping information. The first stage aims to obtain the channel information of each target/eavesdropper. Specifically, the dual-functional BS estimates the direction of each target (potential eavesdropper) via emitting an omnidirectional waveform, which accordingly obtains the estimated target/eavesdropper channel. Then, the secrecy rate and the estimation accuracy can be optimized by formulating the weighted problem iteratively until convergence. Figs.~\ref{two-stage}(b) and \ref{two-stage}(c) illustrate that as the direction of each target/eavesdropper estimation improves, so does the data secrecy rate, revealing a key synergy between sensing and security.

\subsection{Covert Communication in ISAC}
Extending the sensing-aided secure transmission strategies, sensing-assisted covert communication leverages the sensing capabilities of ISAC systems to achieve undetectable communication. This approach capitalizes on the integration of S\&C to design transmission schemes that exploit environmental complexities and adversary behavior for improved covertness.

In \cite{10473676}, by employing extended Kalman filtering (EKF), adversarial movements and channel conditions are accurately tracked, enabling the prediction of target behavior and the design of covert transmission strategies. In deterministic scenarios, the optimization problem is formulated to design radar and communication signals, ensuring undetectability while meeting system constraints. For uncertain multi-path environments, robust fractional programming integrates probabilistic covert constraints, maintaining reliable transmission while obfuscating communication activity. 

This approach transforms environmental challenges into opportunities, enabling ISAC systems to achieve secure and efficient covert communication in adversarial settings.

\section{Open Challenges and Future Work}
\subsection{Passive Target/Eve Classification}
To secure the communication data in ISAC systems, the detection and identification of targets/eavesdroppers poses a significant challenge, particularly when the eavesdropper operates passively. While active eavesdroppers, which transmit signals, can be relatively easily identified and neutralized, passive eavesdroppers remain virtually undetectable due to their silent nature. Moreover, when using ISAC to enable PLS, it is crucial to distinguish passive eavesdroppers from clutter. This creates a substantial vulnerability in ISAC systems, as passive radar systems can intercept confidential communication data without revealing their presence. Addressing this challenge is critical for ensuring the security and integrity of communication data in ISAC environments. Future research must focus on developing advanced techniques and methodologies to prevent passive radar systems from covertly intercepting sensitive information, thereby safeguarding the confidentiality of data in next-generation networks. 
\subsection{Cross-Layer Security}
Current research on network security often focuses on individual layers, such as the physical, MAC, and network layers, addressing security concerns in isolation. However, as we move towards the development of next-generation networks, 6G and beyond, it is increasingly evident that this fragmented approach may be insufficient to address the complex and dynamic security challenges these networks will face. To establish a robust and secure paradigm for 6G and beyond, it is essential to adopt a cross-layer security strategy that integrates and harmonizes security measures across all layers of the network stack. This comprehensive approach is crucial for maintaining the integrity, confidentiality, and availability of data in future networks, where the convergence of communication, sensing, and computing will require unprecedented levels of security coordination across multiple layers.
\subsection{Sensing Privacy Study Solutions are Still Scarce}
Most current research in ISAC systems primarily focuses on securing communication data, including sensed data that is being transmitted, with relatively few studies addressing the critical issue of sensing privacy. Passive radar sensing from sensing eavesdroppers can reveal sensitive information about environments, locations, and even individuals, posing significant privacy risks if not adequately secured. Despite the importance of this issue, research in radar and sensing information privacy remains far from sufficient. To ensure the security and privacy of future networks, it is imperative that future work prioritizes the development of robust frameworks and methodologies to safeguard sensing data, thereby preventing unauthorized access and exploitation. Expanding research in this area will be essential to meet the evolving demands of next-generation ISAC systems and to protect users' privacy in increasingly complex and data-rich environments.

\section{Conclusions}
While ISAC systems offer significant efficiency and performance benefits by integrating communication and sensing functionalities, they also introduce new security challenges that traditional methods fail to address. By exploring advanced PLS techniques, such as artificial noise injection, constructive interference, and sensing-secure transmission, this article has provided a comprehensive overview of potential solutions to eavesdropping and privacy leakage risks. As ISAC systems are further implemented in future networks, ongoing research into passive eavesdropper detection, cross-layer security strategies, and privacy-preserving mechanisms will be essential to ensure both efficient and secure deployments.

\section{Acknowledgement}
N. Su would like to acknowledge the financial support of the NSFC under Grant 62401181 and the Research Matching Grant Scheme from the Research Grants Council of Hong Kong. F. Liu would like to acknowledge the financial support of the Guangdong Basic and Applied Basic Research Foundation under Grant 2024A1515011218. The works of C. Masouros and G. C. Alexandropoulos have been supported by the Smart Networks and Services Joint Undertaking (SNS JU) 6G MUSICAL and 6G-DISAC under the EU’s Horizon Europe research and innovation programme under Grant Agreement numbers 101139176 and 101139130, respectively. The work of Q. Zhang has been supported by the Major Key Project of PCL of China (No. PCL2024A01) and the NSFC under Grant 62027802. The work of T.-T. Chan was supported in part by the Research Matching Grant Scheme from the Research Grants Council of Hong Kong.


 




\ifCLASSOPTIONcaptionsoff
  \newpage
\fi



\bibliographystyle{IEEEtran}
\bibliography{IEEEabrv,CEP_REF}
\end{document}